# Background for Terrestrial Antineutrino Investigations: Radionuclide Distribution, Georeactor Fission Events, and Boundary Conditions on Fission Power Production

**J. Marvin Herndon**

and

**Dennis A. Edgerley**

Transdyne Corporation
San Diego, California 92131 USA

January 24, 2005

**Abstract**

Estimated masses of fissioning and non-fissioning radioactive elements and their respective distributions within the Earth are presented, based upon the fundamental identity of the components of the interior 82% of the Earth, the endo-Earth, with corresponding components of the Abee enstatite chondrite meteorite. Within limits of existing data, the following generalizations concerning the endo-Earth radionuclides can be made: (1) Most of the $^{40}K$ may be expected to exist in combination with oxygen in the silicates of the lower mantle, perhaps being confined to the upper region of the lower mantle where it transitions to the upper mantle; (2) Uranium may be expected to exist at the center of the Earth where it may undergo self-sustaining nuclear fission chain reactions, but there is a possibility some non-fissioning uranium may be found scattered diffusely within the core floaters which are composed of CaS and MgS; and, (3) Thorium may be expected to occur within the core floaters at the core-mantle boundary, although its presence as well at the center of the Earth cannot be ruled out. Results of nuclear georeactor numerical simulations show: (1) The maximum constant nuclear fission power level is 30 terawatts; (2) U-235 comprises 76 percent of present-day georeactor fission, U-238 comprises 23 percent; and, (3) Thorium can neither be fuel nor converted into fuel for the georeactor.

Key Words: Earth Composition, Earth Core, Georeactor, Antineutrino, Nuclear Fission

Communications: mherndon@san.rr.com     http://NuclearPlanet.com

## Introduction

Nuclear fission produces neutron-rich fragments which subsequently $\bar{\beta}$ decay, yielding antineutrinos. Investigations of antineutrinos emanating from within the Earth offer the possibility of validating Herndon's concept of a Terracentric natural nuclear reactor (de Meijer et al. 2004; Domogatski et al. 2004; Fiorentini et al. 2004; Raghavan 2002).

Since 1940, geophysics has progressed under the erroneous assumption that the internal composition of the Earth is like that of an ordinary chondrite meteorite, which has led to widespread misunderstanding that the Earth's inventory of naturally occurring radionuclides is confined to the mantle and the crust. By contrast, Herndon has shown that inner 82% of the Earth's mass is not like an ordinary chondrite at all, but is instead like one of the considerably less-oxidized enstatite chondrites (Herndon 1979; Herndon 1980; Herndon 1982). The consequence of the Earth being less oxygen-rich than previously thought has fundamentally different geophysical implications (Herndon 1998). These include the expected occurrence of radionuclides in the core and the feasibility of a naturally occurring nuclear reactor, the georeactor, at the center of the Earth as the energy source for the planetary geomagnetic field (Herndon 1993; Herndon 1994; Herndon 2003).

To design accurate instrumentation to detect and register geo-antineutrinos with spectral and directional resolution, information is needed on the estimated masses of fissioning and non-fissioning radioactive elements and their respective distributions within the Earth. The purpose of this report is to provide that, to disclose the relative proportions of georeactor fissioning nuclides, and to provide some boundary conditions on georeactor output power along with a brief discussion of the limitations to present knowledge in these areas. Herndon has already given the historical background and discussed the unsuitability of currently popular models that are based upon arbitrary assumptions (Herndon 2004).

## Location of Radionuclides in the Endo-Earth

The Earth consists in the main of two distinct reservoirs of matter separated by the seismic discontinuity that occurs at a depth of about 680 km which separates the mantle into upper and lower parts (Herndon 1980). The endo-Earth, the inner 82% of the Earth's mass consists of the highly reduced lower mantle and core; the more oxidized exo-Earth is comprised of the components of the upper mantle and crust.

The seismically-deduced structure, divisions, and components of the endo-Earth are essentially identical to corresponding parts of the Abee enstatite chondrite meteorite, as shown by the mass ratio relationships presented in Table 1. The identity of the components of the Abee enstatite chondrite with corresponding components of the Earth means that with reasonable confidence one can understand the composition of the Earth's core by understanding the Abee meteorite or one like it. High pressures, such as prevail within the Earth's core, cannot change the state of oxidation of the core. The oxidation state determines not only the relative mass of the core, but the elements the core contains. Highly reduced matter, like that of the Abee enstatite chondrite and the endo-Earth, was separated from primordial solar gases under conditions that severely limited the oxygen content (Herndon & Suess 1976). As a consequence, certain elements, including Si, Mg, Ca, Ti, U, and Th, which would occur entirely as silicate-oxides in ordinary chondrites, occur in part in the iron-based alloy portion of the Abee enstatite chondrite and in the Earth's core; upon cooling these ultimately precipitate as non-oxides, mainly as sulfides.

For decades there has been much discussion as to the possible existence of $^{40}$K in the Earth's core. Although there are some indications from enstatite meteorites of alloy-originated potassium in the mineral djerfisherite (Fuchs 1966), the relative proportion of non-oxide potassium appears to represent at most only a few percent of the potassium. In the Abee enstatite chondrite, most of the potassium occurs in the mineral plagioclase, which would seem to suggest that most of the endo-Earth's $^{40}$K occurs in the lower mantle, perhaps in the region near the boundary of the upper mantle. Additional investigations are needed to be any more precise regarding the distribution of $^{40}$K.

Although there may be some intrinsic uncertainty as to amount of $^{40}$K, if any, in the Earth's core, current data on the uranium distribution in enstatite chondrites clearly indicate the non-lithophile behavior of that element in EH or E4 enstatite chondrites like the Abee meteorite and by inference in the endo-Earth. Generally, uranium occurs within the mineral oldhamite (CaS), an indication that in the enstatite chondrite matter, uranium is a high temperature precipitate. Chemical leaching experiments show that Abee uranium behaves as a sulfide (Matsuda et al. 1972). The tentative assignment of uranium as US or $US_2$ seems reasonable. Regrettably, despite the availability of instrumentation with the capability for determining this information quite precisely, investigators have thus far neglected to do so.

Within the Earth's core, one would expect uranium to precipitate at a high temperature. Just as uranium, a trace element, was swept-up or co-precipitated with a more abundant high temperature precipitate (CaS) in enstatite chondrites, one might expect to some extent the possibility of a similar fate within the Earth's core. Ultimately, uranium, being the densest substance, would be expected to collect at the Earth's center. Unlike other trace elements, such as thorium, uranium nodules on the order of 100 kg early in Earth's history would have been able to maintain sustained nuclear fission chain reactions that could generate sufficient heat to melt their way out of any mineral-occlusion impediment on their descent to the center of the Earth.

Like uranium, thorium occurs exclusively in the alloy portion of the Abee enstatite chondrite and by implication in the Earth's core. Also like uranium, thorium occurs in that meteorite within the mineral oldhamite, CaS (Murrell & Burnett 1982), an indication of its being a high temperature precipitate. Chemical leaching experiments indicate that the Abee thorium behaves in part as a sulfide, and in part as an unknown non-sulfide (Matsuda et al. 1972). Unlike uranium, accumulations of thorium would not have been able to sustain nuclear fission chain reactions.

From the above statements, it would appear that uranium and thorium may occur at the core-mantle boundary occluded in the core floaters, the low density, high temperature CaS and MgS atop the fluid core or, alternatively, they may be concentrated at the center of the Earth, depending upon respective precipitation and accumulation dynamics. Presently, there is no methodology by which to predict the relative proportion of these at the two boundaries of the core, its center and its surface. Because of the ability of 100 kg nodules of uranium to undergo self-sustaining nuclear fission chain reactions, which can melt free of occlusion, one might expect appreciable uranium to occur at the center of the Earth. By contrast, from the mineral data on the Abee meteorite, one might expect most of the potassium to occur near the upper boundary of the lower mantle, with the possibility of only a small amount occurring in the fluid core. Quantitative estimates of present-day total amounts of these radionuclides are shown in Table 2; the boundary values shown for uranium are the range between no fission and maximum (constant level) fission.

The remaining 18% of the Earth's mass, the exo-Earth, consists of the upper mantle plus crust. From seismic discontinuities, from the observed FeO content of deep-source rocks (indicative of oxidation state), and from the existence of at least one undifferentiated chondritic component, the exo-Earth appears to be more oxidized than the endo-Earth and appears to consist of layers of veneer.

## Nuclear Fission at the Center of the Earth

Unlike thorium, uranium nodules greater than about 100 kg can undergo self-sustaining nuclear fission chain reactions. Nuclear fission produces energy, consumes uranium, and produces neutron-rich fission products which subsequently β decay, yielding antineutrinos. In demonstrating the feasibility of a Terracentric nuclear reactor, Herndon used very conservative uranium estimates, amounting to approximately 20% of the total initial endo-Earth uranium content shown in Table 2. In the present paper, the results of numerical simulations are presented which provide some boundary conditions assuming that the entire amount of uranium is available for fission. The specific relative proportions of fissioning nuclides are determined. Ancillary calculations show that under deep-Earth georeactor conditions, thorium will not be fertile and thus will be unable to contribute significantly toward georeactor fuel breeding.

Numerical simulation calculations were made using the SAS2 analysis sequence contained in the SCALE Code Package from Oak Ridge National Laboratory (SCALE 1995) that has been developed over a period of three decades and has been extensively validated against isotopic analyses of commercial reactor fuels (DeHart & Hermann 1996; England et al. 1984; Hermann 2000; Hermann et al. 1995; Hermann & DeHart 1998). The SAS2 sequence invokes the ORIGEN-S isotopic generation and depletion code to calculate concentrations of actinides, fission products, and activation products simultaneously generated through fission, neutron absorption, and radioactive decay. The SAS2 sequence performs the 1-D transport analyses at selected time intervals, calculating an energy flux spectrum, updating the time-dependent weighted cross-sections for the depletion analysis, and calculating the neutron multiplication of the system. As in previous numerical simulations (Herndon 2003; Hollenbach & Herndon 2001), time steps of 2 X $10^6$ years were used throughout, nuclear georeactor operation was assumed to have commenced 4.5 X $10^9$ years ago and ceased when the effective neutron multiplication constant $K_{eff}$ < 1, and in all cases fission products were promptly removed upon formation; 36.84 g/cm$^3$ was used for uranium density with initial uranium amounts shown in Table 3.

Calculations were made at different steady-state fission power levels to ascertain the highest constant power level at which the georeactor could operate and still be functioning after 4.5 X $10^9$ years. The maximum constant power level was determined to be 30 TW (terawatts) for the Terracentric georeactor using the maximum uranium content at 4.5 X $10^9$ years ago, as shown in Table 2. Present-day radionuclide abundance estimates are presented in Table 3. For reference, calculations were also made at 4 TW using the smaller amount of uranium that was used in previous numerical simulations (Herndon 2003; Hollenbach & Herndon 2001). The proportions of major and some minor fissioning nuclides as a function of time are shown in Figs. 2 and 3, respectively. The present-day numerical values for the fractions of these and some even less abundant fissioning nuclides are shown for the two cases in Table 4. Notably, there are only two dominant fissioning nuclides: $^{235}$U (76%) and $^{238}$U (23%). As noted by Herndon (1994) and by Seifritz (2003), the principal fuel-breeding takes place by the reaction

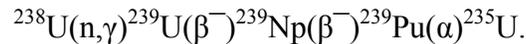

$$^{238}U(n,\gamma)^{239}U(\beta^-)^{239}Np(\beta^-)^{239}Pu(\alpha)^{235}U.$$

Much is yet unknown concerning the distribution of radionuclides within the Earth. Because of the identity between the parts of the endo-Earth and corresponding parts of the Abee enstatite chondrite, it is possible to make direct inferences as to radionuclide states of oxidation and locations within the endo-Earth, although not to the degree of precision that might ultimately be possible given adequate petrologic data with modern instrumentation and appropriate laboratory experiments (Herndon 1996; Herndon 1998; Herndon 2004).

Within those limitations, the following generalizations concerning the endo-Earth radionuclides can be made: (1) Most of the $^{40}$K may be expected to exist in combination with oxygen in the silicates of the lower mantle, perhaps being confined to the upper region of the lower mantle where it transitions to the upper mantle; (2) Uranium may be expected to exist at the center of the Earth where it may undergo self-sustaining nuclear fission chain reactions, but there is a possibility some non-fissioning uranium may be found scattered diffusely within the core floaters which are composed of CaS and MgS; and, (3) Thorium may be expected to occur within the core floaters at the core-mantle boundary, although its presence as well at the center of the Earth cannot be ruled out. Thorium is unable to be fuel for or to be converted into fuel for the georeactor.

It would be desirable to be able to specify the radionuclide distribution within the exo-Earth, the upper mantle and crust. But at present there is uncertainty in the compositions of the layers of the upper mantle and uncertainty as to the composition of the parent materials for that region of the Earth. As a "ball park" estimate, one might guess that the radionuclide complement of the exo-Earth represents an additional 22% of the endo-Earth complement, with much of the exo-Earth uranium and thorium residing in the crust. Ultimately, it should be possible to refine these estimates, not by making models based upon arbitrary assumptions, but by tedious efforts to discover fundamental quantitative relationships that lead logically to that information. Endo-Earth radionuclide estimates and exo-Earth radionuclide guesses are shown in Table 2.

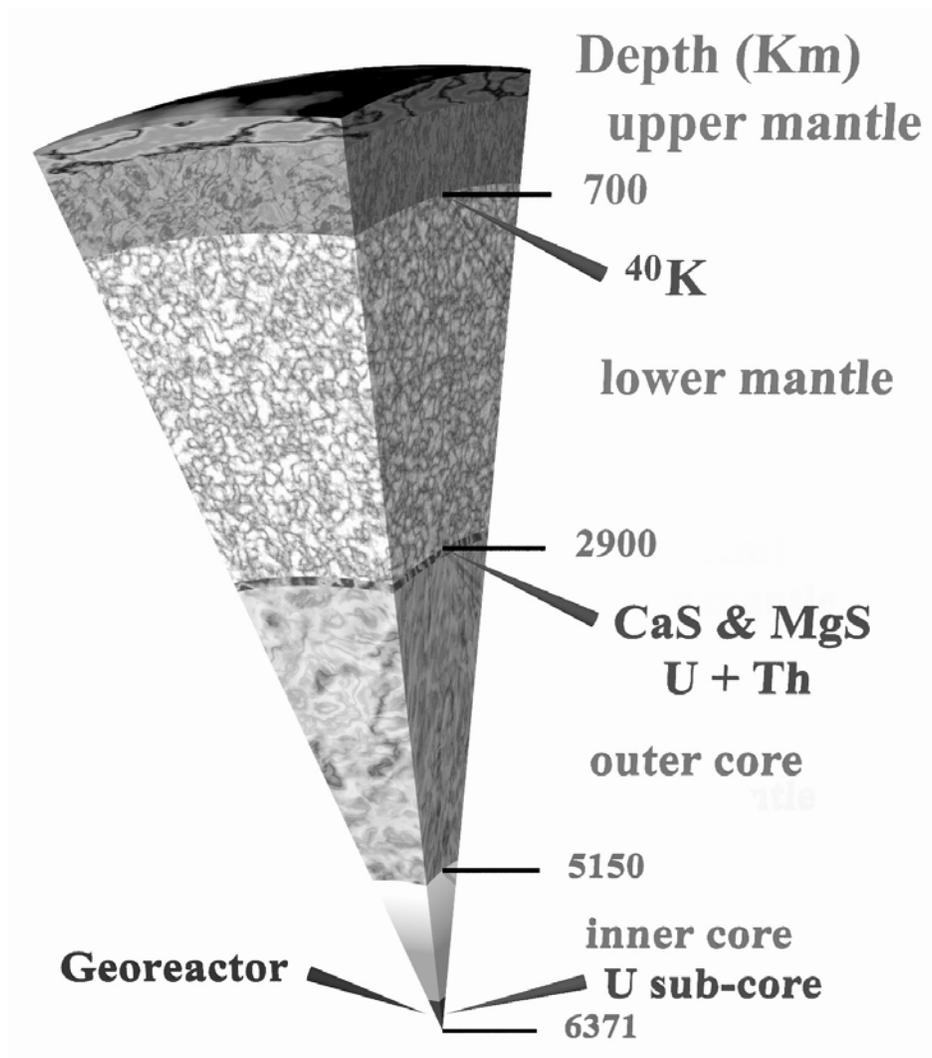

**Figure 1.** Schematic representation of the interior of the Earth showing regions in the endo-Earth where radionuclides may be expect to be concentrated.

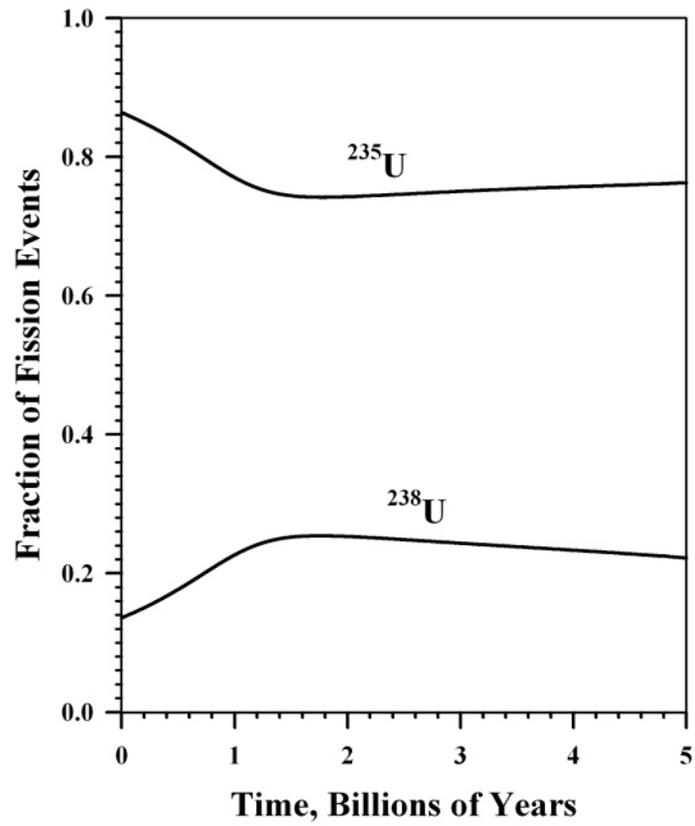

**Figure 2.** Fraction of fission events of the major fissioning species as a function of time in the 30 TW numerical simulation. Similar results were obtained in the 4 TW simulation.

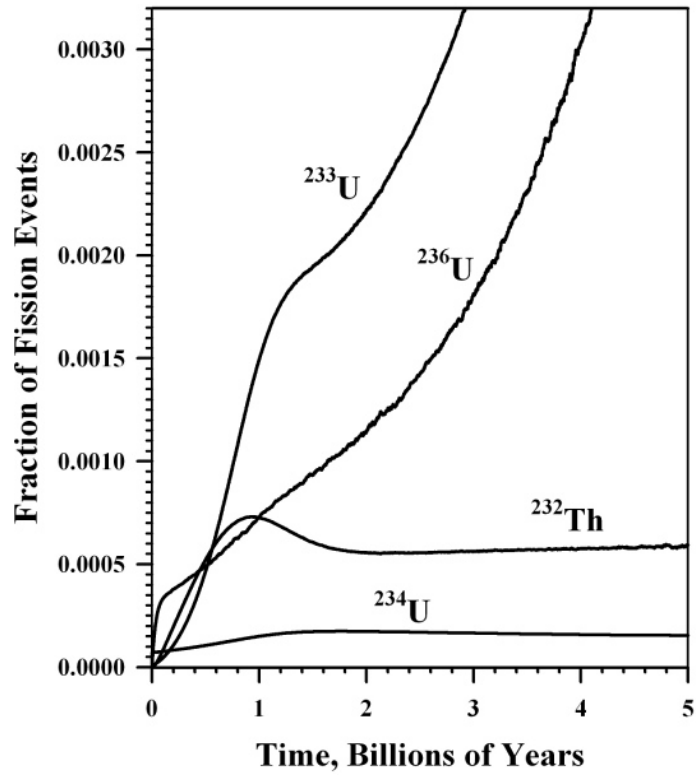

**Figure 3.** Fraction of fission events of the more abundant minor fissioning species as a function of time in the 30 TW numerical simulation. Similar results were obtained in the 4 TW simulation.

**Table 1**. Fundamental mass ratio comparison between the endo-Earth (core plus lower mantle) and the Abee enstatite chondrite. From (Herndon 1980; Herndon 1982; Herndon 1993).

| Fundamental Earth Ratio | Earth Ratio Value | Abee Ratio Value |
|---|---|---|
| lower mantle mass to total core mass | 1.49 | 1.43 |
| inner core mass to total core mass | 0.052 | *theoretical* 0.052 if $Ni_3Si$ 0.057 if $Ni_2Si$ |
| inner core mass to (lower mantle+core) mass | 0.021 | 0.021 |
| core mean atomic mass | 48 | 48 |
| core mean atomic number | 23 | 23 |

**Table 2**. Estimates of the maximum present-day radionuclide content within the endo-Earth and guessed amounts in the exo-Earth. Endo-Earth values of uranium in parentheses, given for reference only, assume no fission activity. Data from (Baedecker & Wasson 1975; Murrell & Burnett 1982)

| Endo-Earth (estimate) | | |
|---|---|---|
| *Nuclide* | *Kilograms* | |
| $^{40}$K | $5.001 \times 10^{17}$ | |
| $^{232}$Th | $1.322 \times 10^{17}$ | |
| $^{235}$U | $3.065 \times 10^{14}$ | $(2.504 \times 10^{14})$ |
| $^{238}$U | $3.373 \times 10^{15}$ | $(3.456 \times 10^{16})$ |

| Exo-Earth (guess) | |
|---|---|
| *Nuclide* | *Kilograms* |
| $^{40}$K | $1.100 \times 10^{17}$ |
| $^{232}$Th | $2.908 \times 10^{16}$ |
| $^{235}$U | $4.629 \times 10^{15}$ |
| $^{238}$U | $1.528 \times 10^{16}$ |

**Table 3.** Initial uranium parameters used in the two numerical simulations.

| Initial Parameters | 4 TW Simulation | 30 TW Simulation |
|---|---|---|
| *Nuclide* | *Kilograms* | |
| $^{235}$U | 4.8672 X $10^{15}$ | 2.1039 X $10^{16}$ |
| $^{238}$U | 1.6060 X $10^{16}$ | 6.9427 X $10^{16}$ |

**Table 4**. Present-day fraction of fission events for the two numerical simulations.

| Nuclide | 4 TW Fraction of Fission Events at Present Time | 30 TW Fraction of Fission Events at Present Time |
|---|---|---|
| $^{235}$U | 7.46E-01 | 7.60E-01 |
| $^{238}$U | 2.49E-01 | 2.28E-01 |
| $^{233}$U | 2.60E-03 | 6.90E-03 |
| $^{236}$U | 1.38E-03 | 3.97E-03 |
| $^{232}$Th | 5.63E-04 | 5.77E-04 |
| $^{239}$Pu | 1.70E-04 | 3.53E-04 |
| $^{237}$Np | 1.43E-04 | 2.14E-04 |
| $^{229}$Th | 6.96E-05 | 1.84E-04 |
| $^{234}$U | 4.99E-05 | 1.53E-04 |
| $^{231}$Pa | 6.49E-06 | 6.99E-06 |
| $^{232}$U | 2.13E-09 | 6.07E-09 |
| $^{240}$Pu | 2.49E-10 | 1.36E-09 |
| $^{238}$Pu | 1.99E-10 | 3.47E-10 |
| $^{227}$Th | 3.08E-11 | 2.57E-10 |
| $^{237}$U | 5.30E-13 | 2.14E-12 |
| $^{238}$Np | 4.72E-14 | 5.20E-13 |
| $^{241}$Am | 5.88E-17 | 1.10E-15 |
| $^{241}$Pu | 1.11E-17 | 1.97E-16 |
| $^{242}$Am | 3.95E-23 | 1.90E-21 |